# An Actuator with Magnetic Restoration, Part II: Drive Circuit and Control Loops

Sajjad Mohammadi, *Member, IEEE*, William R. Benner, James L. Kirtley, *Life Fellow*, Jeffrey H. Lang, *Life Fellow*

*Abstract*—In part II, an op-amp-based drive is proposed and designed. Subsequently, a very accurate model for the drive circuit and the current loop is developed as a simulation platform, while its simplified version is derived, tailored for efficient design purposes. Through a comprehensive evaluation, the accuracy and efficacy of both the actuator and drive circuit modeling is scrutinized, showcasing their superiorities over existing approaches. The importance of eddy current modeling is underscored. Also, the effectiveness of the designed current loop and its practical trade-offs are engineered and discussed. Then, three DSP-based position control techniques are implemented: pole placement with voltage drive, pole placement with current drive, and nonlinear control with feed linearization. Both full-order and reduced-order observers are leveraged to estimate the unmeasured states. The performance of control designs across various applications are evaluated through indices such as rise time, overshoot, steady-state error, and large-signal tracking in the step response as well as bandwidth, robustness, phase margin, sensitivity, disturbance rejection, and noise rejection in the frequency domain. The distinctive features of implemented control strategy are compared, offering a nuanced discussion of their respective advantages and drawbacks, shedding light on their potential applications.

## I. INTRODUCTION

As rotary actuators and voice coil motors (VCMs) continue to play a vital role in various industries, the exploration of these devices has captivated researchers [1]-[3]. Part I of this study introduced an electromechanical model for the actuator, considering the influence of eddy currents and pre-sliding friction. The model underwent evaluation and identification through an experimental prototype. Part II shifts focus to the drive circuit and control loops. In [4], an open-loop control of a VCM with magnetic restoration is detailed, utilizing an electromechanical model that neglects eddy currents in the yoke and magnets while incorporating friction. Closed-loop position control systems for rotary actuators can be achieved through voltage drives [5]-[7] or current drives [8]-[12]. However, existing electromechanical models used in these control systems often overlook model nonlinearities and eddy currents in laminations and magnets, introducing inaccuracies and hindering efficiency in designs.

Voltage drives, while cost-effective and simple, come with drawbacks such as slower response times, limited robustness, and increased model uncertainties. In [5], the implementation of design, modeling, and seek control for a VCM actuator using a voltage drive is explored, in which nonlinearities and eddy currents are neglected. [6] utilizes linear power amplifiers with a bridge circuit to drive actuator coils in a FlexLab/LevLab system, employing a myRIO by National Instruments as a real-time controller. [7] introduces a controller where the conventional current driver for the VCM is replaced by a voltage driver, replicating the behavior of a current loop through a model of the VCM and its driver; however, this developed model overlooks nonlinearities and eddy currents.

Current drives, leveraging a high-bandwidth current loop to eliminate the electrical dynamic of the actuator, provide a faster response, enhanced robustness, and potential model simplifications for advanced controls. In [8], control of a novel levitated hysteresis bearingless slice motor is realized using an analog circuit featuring a linear power op-amp as the power stage and three high-voltage op-amps as the control stage. Position and suspension control algorithms are implemented in a field-programmable gate array (FPGA) of an NI cRIO-9076 target via LabVIEW. [9] employs two linear three-phase transconductance amplifiers to separately drive the suspension and rotation windings as a current control loop for a homopolar bearingless motor, while suspension and torque controllers are implemented by FPGAs of a real-time controller (NI cRIO-9064). In [10], an advanced digital controller is implemented for a direct-drive servo valve with a high-frequency VCM, utilizing an FPGA-based H-bridge switching drive for the current loop and a DSP for the pole placement position controller. The paper [11] develops an electromechanical model for a linear VCM considering frictional load while ignoring eddy currents. Then, direct amplitude control is implemented using a full-bridge inverter. In [12], a linear electromechanical model for a VCM is developed which ignores eddy currents, and two digital current controllers are compared: traditional PID and a robust control based on the disturbance observer framework. Finally, [13] implements an adaptive fuzzy PID controller based on a electrotechnical model, neglecting eddy currents and nonlinearities, showcasing superior performance in nonuniform friction, disturbance variation, and load changes compared to traditional PID controllers.

Feedback linearization nonlinear control is a valuable approach for achieving large signal tracking. In [14], this technique is successfully employed for a magnetic suspension system, utilizing a developed nonlinear electromechanical model. The implementation involves a high-bandwidth current loop and a digital computer. In [15], a feedback linearization nonlinear speed controller is offerd for a permanent magnet synchronous motor (PMSM).

Model-free controls offer an alternative perspective when precise electromechanical models face challenges due to unknown plant dynamics. In [16], an innovative adaptive dynamic sliding-mode fuzzy cerebellar model articulation controller system is introduced for actuators. Additionally, [17] presents an alternative strategy to circumvent the challenges associated with modeling the nonlinearities of VCMs, which is achieved through the utilization of a B-spline neural network for position control.

State observers are crucial in estimating unmeasured states or enabling sensorless control. an implementation of field-oriented control (FOC) for a hysteresis motor is highlighted in [18], employing a flux orientation observer based on a state space model of the motor. Additionally, [19] introduces innovative rotor position and stator current estimators tailored for sensorless direct torque control (DTC) of BLDC motors. The quest for improved performance in the FOC of interior permanent magnet synchronous motors (IPMSM) is addressed in [20] through the development of an advanced model-based high-order sliding-mode observer. Finally, [21] provides a comprehensive analysis of control system designs, elucidating associated trade-offs such as the six gangs.

Part II of the paper shifts its focus to the modeling, design, and evaluation of the drive and different control loops. An op-amp-based drive circuit is proposed, modeled, and designed. Leveraging a third-order op-amp model, a highly accurate model for the drive circuit and the current control loop is developed, providing an excellent simulation and prediction platform. Furthermore, a simplified version of the model is derived, aiding in design considerations and offering insights into the control loop dynamics. The effectiveness of the modeling for both the actuator and the drive circuit is scrutinized across diverse control scenarios. Emphasizing the significance of incorporating nonlinearities and eddy currents in the modeling and design of control loops, it is illustrated how neglecting eddy currents can lead to inaccuracies in phase margin and sensitivity. Additionally, the effectiveness of the designed current loop and various practical trade-offs are discussed.

Afterwards, three DSP-based position control techniques are implemented. Firstly, an exploration into pole placement position control with voltage drive is conducted, demonstrating satisfactory performance for simple applications, while showing limitations for more complex control scenarios due to uncertainties and unmodeled dynamics in the electrical domain. Secondly, leveraging the developed current control loop, an investigation into pole placement position control with a current drive is performed. This approach enhances accuracy, bandwidth, and robustness by eliminating electrical dynamics, including eddy currents, back-emf, and temperature-dependent resistance of the coil. Despite these improvements, its performance diminishes in large-signal control. Thirdly, benefiting from precise actuator modeling, a feedback linearization technique is implemented for nonlinear control, specifically tailored for large-signal tracking. Full-order and reduced-order observers are incorporated to estimate unmeasured states. Evaluation and comparison of control system designs are conducted using indices such as rise time, overshoot, and steady-state error in the step response, along with bandwidth, phase margin, sensitivity, disturbance rejection, and noise rejection in the frequency domain. The implemented controls are juxtaposed, and their respective advantages and disadvantages concerning performance, cost, hardware, and complexity are discussed.

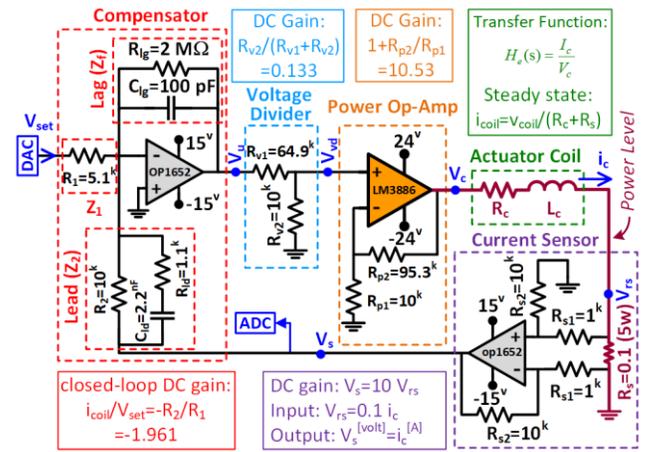

Fig. 1. Drive circuit and current control loop

## II. DRIVE AND CURRENT LOOP

The time constant of the electrical dynamic can be eliminated using a high bandwidth current control loop. Thus, instead of the coil voltage, the current or torque can be commanded directly. Also, the complexities such as fractional dynamics of eddy currents are removed, resulting in simplicity and higher accuracy of position control. In addition, the robustness of the drive is increase by making the system independent of temperature-dependent elements such as coil resistance. The op-amps are not temperature sensitive as long as operating within the published specs. As shown in Fig. 1, an op-amp-based analog drive is employed for the current loop, for which a very accurate model is developed using a non-ideal model of the op-amps. According to the datasheets of the op-amps, a third-order model for the open-loop gain $A(s)$ can be approximated as

$$\begin{cases} A(s) = \dfrac{A_{OL}}{(1+\dfrac{s}{2\pi f_1})(1+\dfrac{s}{2\pi f_2})(1+\dfrac{s}{2\pi f_3})} \\ f_1 \approx \dfrac{GBP}{A_{OL}} \end{cases} \quad (1)$$

where the open-loop gain $A_{oL}$ and gain-bandwidth product $GBP$ are given in the datasheet, and frequencies $f_1$, $f_2$ and $f_3$ are approximated to obtain a 3$^{rd}$ order model. The approximated frequency responses of LM3886 and OP1652 are shown in Fig. 2. By writing the differential input voltage $V_d$ in terms of output voltage $V_o$ and inputs $V_+$ and $V_-$, the op-amp is modeled:

$$\begin{cases} V_o = A(s)V_d \\ V_d = V_+ - V_- \end{cases} \quad (2)$$

The developed model is very precise for simulations. However, for the design procedure, a simplified model in terms of conventional control systems is obtained with infinite gain $A(s)$. The block diagrams of both models are given in Fig. 3.

### A. Modeling of the Power Op-Amp and Voltage Divider

The power op-amp LM3886 with an open-loop gain of $A_1(s)$ can deliver a current of ±10 A. Generally, since the gain-bandwidth product (GBP) of the op-amp is a fixed value, the higher the closed-loop gain, the lower the bandwidth; thus, the lowest possible gain is preferred. It is visualized in Fig. 2(a). Based on the datasheet of LM3886, there is a lowest closed-loop gain to have a stable circuit with enough phase margin. Therefore, mid-range values for $R_{p1}$ and $R_{p2}$ are picked to have a gain of 10.53. A voltage divider with a gain of 0.133 is used to adjust the maximum output of the compensator (±14.7) to the maximum output of the power op-amp (±14.7 volt×0.133×10.53=±20.6 volt). The transfer function of the voltage divider is just a gain as

$$H_{vd}(s) = \frac{V_{vd}}{V_u} = \frac{R_{v2}}{R_{v1}+R_{v2}} \quad (3)$$

The differential input voltage of the power op-amp is as

$$\begin{cases} v_{dp} = V_{vd} - \dfrac{R_{p1}}{R_{p1}+R_{p2}}V_c \\ V_c = A_1(s)v_{dp} \end{cases} \quad (4)$$

The transfer function of the ideal model is just a gain as

$$\begin{cases} H_p(s) = \dfrac{V_c}{V_{vd}} = \dfrac{A_1(s)}{1+\dfrac{R_{p1}}{R_{p1}+R_{p2}}A_1(s)} \\ \lim_{A_1(s)\to\infty} H_p(s) = 1+\dfrac{R_{p2}}{R_{p1}} \end{cases} \quad (5)$$

Its bandwidth is large enough compared to the current loop bandwidth that can be treated as a gain in the design process.

### B. Modeling of the Current Sensor

A low-noise high-bandwidth op-amp OP1652 with an open-loop gain of $A_2(s)$ is used for compensator and current measurement. The coil current is measured by the voltage across a Metal Element 5-watt resistance $R_s$ =0.1 Ω in series with the coil, whose voltage is buffered so that it is not loaded. This open-air resistor keeps the hot spot safely off the PCB and improves heat dissipation. Also, the sense circuitry should be as close as possible to the sense resistor to avoid large loop areas by the PCB tracks, which can form parasitic inductances. The buffer gain is set to $1/R_s$, i.e., $R_{s2}/R_{s1}$=10, so the DC gain of $H_s$ is unity ($V_s=i_c$). The differential input voltage of the op-amp is

$$
\begin{cases}
v_{ds} = \dfrac{R_{s1}}{R_{s1}+R_{s2}}\left(V_{rs}-\dfrac{R_{s1}}{R_{s2}}V_s\right) \\
V_{rs} = R_s i_c \; ; \;\; V_s = A_2(s)v_{ds}
\end{cases}
\tag{6}
$$

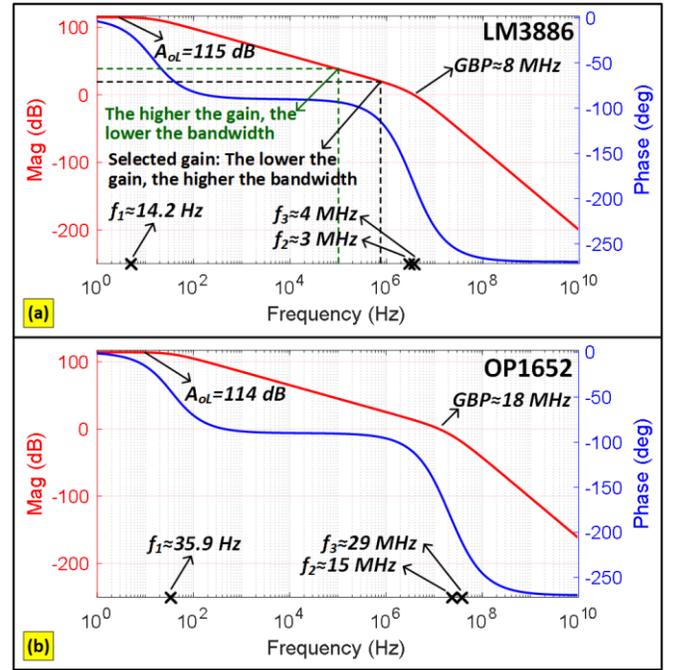

Fig. 2. Approximated frequency responses of (a) LM3886 and (b) OP1652

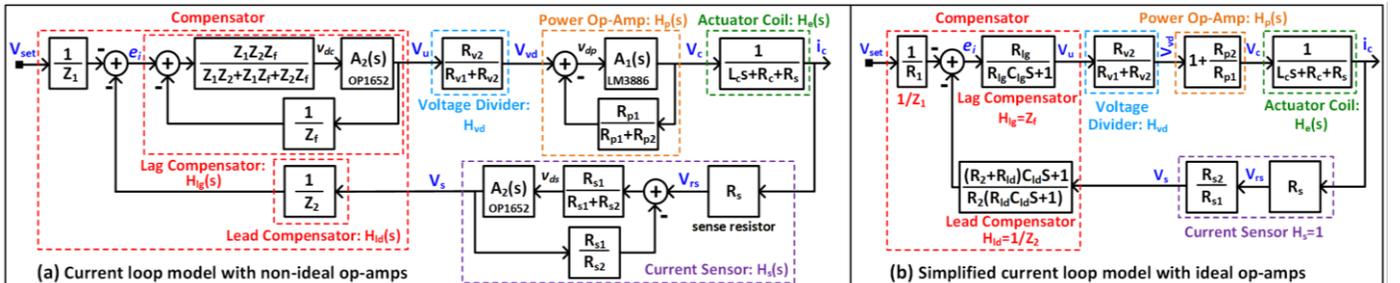

Fig. 3. Modeling of the drive circuit using (a) nonideal op-amp models for simulation, and (b) ideal op-amp models for design of the current control loop.

The transfer function of the ideal model is just a gain as

$$
\begin{cases}
H_s(s) = \dfrac{V_s}{I_c} = R_s \dfrac{\dfrac{R_{s1}}{R_{s1}+R_{s2}}A_2(s)}{1+\dfrac{R_{s1}}{R_{s1}+R_{s2}}\dfrac{R_{s1}}{R_{s2}}A_2(s)} \\
\lim\limits_{A_2(s)\to\infty} H_s(s) = R_s \dfrac{R_{s2}}{R_{s1}} = 1
\end{cases}
\tag{7}
$$

It is treated as a gain in the design of the current loop.

### C. Modeling of the Lead-Lag Compensator

The lag compensator provides a large low-frequency gain to eliminate steady-state error. The lead compensator provides a fairly large phase margin to limit the overshoot of the time response and to increase the robustness of the control system.

The lead compensator is placed in the feedback path to reduce overshoot in the output of the power op-amp. The differential input of the op-amp $V_{dc}$ is as

$$
-v_{dc} = \dfrac{Z_f\parallel Z_2}{Z_1+Z_f\parallel Z_2}V_{set}+\dfrac{Z_f\parallel Z_1}{Z_2+Z_f\parallel Z_1}V_s+\dfrac{Z_1\parallel Z_2}{Z_f+Z_1\parallel Z_2}V_u
\tag{8}
$$

It can be simplified to:

$$
\begin{cases}
v_{dc} = -\dfrac{Z_1 Z_2 Z_f}{Z_1 Z_2 + Z_1 Z_f + Z_2 Z_f}\left\{\dfrac{V_{set}}{Z_1}+\dfrac{V_s}{Z_2}+\dfrac{V_u}{Z_f}\right\} \\
V_u = A_2(s)v_{dc}
\end{cases}
\tag{9}
$$

The transfer function of the lag compensator is obtained as

$$
H_{lg} = \dfrac{V_u}{e_i} = \dfrac{\dfrac{Z_1 Z_2 Z_f}{Z_1 Z_2 + Z_1 Z_f + Z_2 Z_f}A_2(s)}{1+\dfrac{Z_1 Z_2 Z_f}{Z_1 Z_2 + Z_1 Z_f + Z_2 Z_f}\dfrac{1}{Z_f}A_2(s)}
\tag{10}
$$

With the ideal model of op-amps ($A_2(s)\to\infty$), it reduces to

$$
\begin{cases}
H_{lg}(s) = Z_f = \dfrac{R_{lg}}{R_{lg}C_{lg}s+1} \\
\lim\limits_{R_{lg}\to\infty} H_{lg}(s) = \dfrac{1}{C_{lg}s}
\end{cases}
\tag{11}
$$

The large resistor $R_{lg}$ limits the DC gain of the loop to avoid overcurrent in the coil in unexpected scenarios. If $R_{lg}$ is very large as here, a pure integrator is obtained as $1/C_{lg}S$. The simplified transfer function of the lead compensator is

$$
\begin{cases}
H_{ld}(s) = \dfrac{1}{Z_2} = \dfrac{1}{R_2}\dfrac{(R_2+R_{ld})C_{ld}s+1}{R_{ld}C_{ld}s+1} \\
H_{ld}(s) = \dfrac{\alpha}{R_2}\dfrac{\tau s+1}{\tau s+1}
\end{cases}
\tag{12}
$$

where time constant and pole-zero ratio are $\tau=R_{ld}C_{ld}$, and

pole- $\alpha=1+R_2/R_{ld}$. The lead compensator provides a maximum phase of $\phi_m$ at the frequency of $\omega_m$ as

$$
\phi_m = \sin^{-1}\left(\dfrac{\alpha-1}{\alpha+1}\right) \; at \;\; \omega_m = \dfrac{1}{\tau\sqrt{\alpha}}
\tag{13}
$$

Too big values of $\alpha$ can amplify high-frequency noise. The value of $\omega_m$ is set at the gain crossover frequency $\omega_c$ $=2\pi f_c$ of the loop so that the highest phase margin is obtained.

### D. Design of Lead-Lag Compensator

The closed-loop DC gain is almost $R_2/R_1$, whose value is picked such that bounds of $V_{set}$ ($\pm$5V from DAC of DSP) are matched to the current capability of the power op-amp ($\pm$5$\times$10/5.1=$\pm$9.8A). The resistor $R_1$ should not be smaller than 1 k$\Omega$ to avoid heating and damaging the DAC by drawing a large current. Picking $R_1$=5.1 k$\Omega$, leaves $R_2$=10 k$\Omega$. Next, using a pole-zero ratio of $\alpha$=10, a maximum phase of $\phi_m\approx55^o$ is added to the loop. Therefore, $R_{ld}$ is obtained as

$$
\alpha = 1+\dfrac{R_2}{R_{ld}} \;\; \Rightarrow \;\; R_{ld} = \dfrac{R_2}{\alpha-1} \approx 1.1 \; k\Omega
\tag{14}
$$

The 10%-90% rise time of the closed-loop response is $t_r\approx2.2/\omega_{bw}$, where $\omega_{bw}$ is the bandwidth in rad/sec. At least a bandwidth of 7 kHz is required to have a $t_r$<50$\mu$s. A crossover frequency of $f_c$=20 kHz, which is much larger than $1/\tau_e$, is picked to provide a closed-loop bandwidth around $f_{bw}$=7.8 kHz. Setting $\omega_m=\omega_c=2\pi f_c$, the value of $C_{ld}$ is obtained as

$$
\omega_m = \dfrac{1}{R_{ld}C_{ld}\sqrt{\alpha}}
$$
$$
\Rightarrow \; C_{ld} = \dfrac{1}{\omega_m R_{ld}\sqrt{\alpha}} \approx 2.2 \; nF
\tag{15}
$$

The last component to be determined is $C_{lg}$ which is set such that the gain of loop transmission is unity at $\omega_c$.

$$
\left|\dfrac{1}{j\omega_c C_{lg}}H_{ld}(j\omega_c)H_s(j\omega_c)\dfrac{R_{s2}}{R_{s1}+R_{s2}}\left(1+\dfrac{R_{p2}}{R_{p1}}\right)\right|=1
\tag{16}
$$
$$
\Rightarrow C_{lg} \approx 100\,pF
$$

The electrical dynamic $H_e$ includes eddy currents.

## III. Evaluation and Trade-Offs of Current Loop

Fig. 4 shows the loop transmission and its components. The results of the developed model are in very close agreement with the experiment. A sufficient phase margin of $\phi_m$=72.5$^o$ is obtained. It is seen that the phase margin is estimated with an error less than 1$^o$ with the electrical dynamic including eddy current, while an error of around 16$^o$ is observed if eddy currents are ignored which can be very misleading in the control system design.

Also, the design trade-off of the current control loop is studied. As shown in the block diagram given in Fig. 5, the three important inputs of the current loop are reference $R$ (current command $V_{set}$), disturbance $D$, and measurement noise $N$. Also, the three outputs are the plant output $x$ (position $\theta$), measured output $y$, and drive output $u$. The loop transmission is $L=PCH$. It can be represented as a MIMO system as

$$\begin{bmatrix} x \\ y \\ u \end{bmatrix} = \begin{bmatrix} \dfrac{P}{1+PCH} & \dfrac{-PCH}{1+PCH} & \dfrac{PCF}{1+PCH} \\ \dfrac{P}{1+PCH} & \dfrac{1}{1+PCH} & \dfrac{PCF}{1+PCH} \\ \dfrac{PCH}{1+PCH} & \dfrac{CH}{1+PCH} & \dfrac{CF}{1+PCH} \end{bmatrix} \begin{bmatrix} D \\ N \\ R \end{bmatrix} \quad (17)$$

There are six district transfer functions known as the six gangs [16]. The experimental frequency responses are obtained by SR785 Digital Signal Analyzer, whose maximum frequency is 100 kHz. By obtaining the frequency responses of loop transmission $L=PCH$, Gang 1, Gang 2 and Plant $P$, the Gangs 3-6 can directly be obtained as $G3=P/(1+L)$, $G4=1/(1+L)$, $G5=(L/P)/(1+L)$ and $G6=L/(1+L)$. The high precision of the developed models for the actuator and the drive circuit is illustrated in comparison with the experimental data. It is also shown that the RL model of the electrical dynamic in which the eddy currents are ignored may cause misleading inaccuracies in the design process.

### A. Gang 1: Reference Tracking

This is the reference tracking transfer function from the current command (DAC) to the coil current as in below:

$$\begin{cases} T = \dfrac{Y}{R} = \dfrac{FPC}{1+PCH} \\ T_{DC} = \lim_{C \to \infty} T(0) = \dfrac{F(0)}{H(0)} = \dfrac{R_2}{R_1} \approx 5.96 \, dB \end{cases} \quad (18)$$

Frequency and step responses of $T$ are shown in Figs. 6 and 7. Provided by the crossover frequency of $\omega_c=20$ kHz, as $\omega_{bw} \propto \omega_c$, a sufficient bandwidth of around $f_{bw}=7.86$ kHz is obtained, which provides a fast response with a small rise time $t_r \approx 2.2/\omega_{bw}=45$ us as expected. Also, the bandwidth is not excessively large us to introduce high-frequency noise to the system. Thanks to the sufficient phase margin of the loop, the closed-loop response is well damped ($\zeta \approx \varphi_{pm}/100$) without a significant resonance peak. Provided by the lag compensator, if the loop gain at low frequency is large enough, the steady-state error converges to zero, and the DC gain is $R_2/R_1=1.961$, that is, a current command of $V_{set}=1$ produces a current of 1.961 A in the coil.

### B. Gang 2: Voltage Capability of the Drive

This is the transfer function from the current setpoint $R$ to the output of the power op-amp $U$ as

$$\begin{cases} \dfrac{U}{R} = \dfrac{FC}{1+PCH} \\ \lim_{C \to \infty} \dfrac{FC}{1+PCH} = \dfrac{F(0)}{P(0)H(0)} = \dfrac{R_2}{R_1}(R_c+R_s) \approx 11.26 dB \end{cases} \quad (19)$$

A design criterion is the DC gain which converts the current setpoint (DAC voltage) to the steady-state coil voltage. The DC gain of 11.26 dB converts the ±5 volt at the DAC to ±18 volt at the coil terminal—a bit below the maximum voltage capability of drive. Also, a comparison is made with a case where the lead compensator is placed in the forward path. As shown in Fig. 6, it can be observed that the resonance peak of the frequency response and the overshoot of the step response is larger if the lead compensator is placed in the forward path, which can result in saturation of the power op-amp whose output voltage cannot go beyond ±20.6 volts. Therefore, putting the lead compensator in the feedback path is a wise design that enhances the voltage capability of the drive in the transient regime. The step response is also shown in Fig. 7.

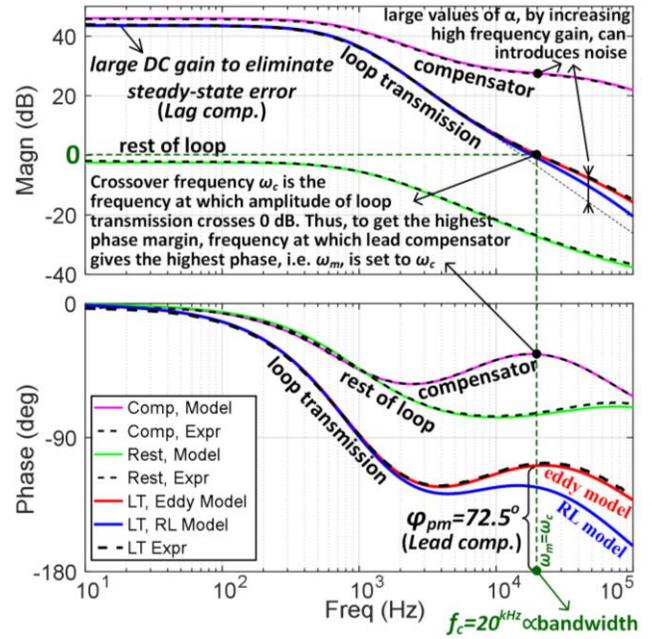

Fig. 4. Frequency response of loop components: loop transmission, compensator, and rest of loop (loop transmission excluding compensator)

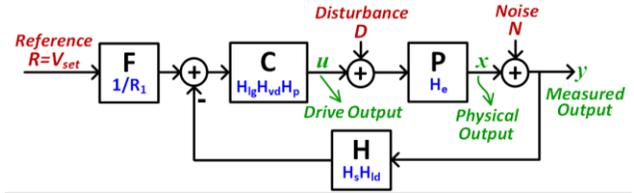

Fig. 5. The six gangs: block diagram, inputs and outputs

### C. Gang 3: Disturbance Rejection or Load Sensitivity

This is the transfer function from the disturbance $D$ to the output $y$ (coil current) as

$$\dfrac{Y}{D} = \dfrac{P}{1+PCH}; \quad \lim_{C \to \infty} \dfrac{P}{1+PCH} = 0 \quad (20)$$

The disturbance operates at low frequency as the reference command. The back-emf $E=k_b\omega_r$ is treated as a disturbance in the current loop. A large loop gain in low frequencies provided by the lag compensator brings a good disturbance rejection whose capability needs a compromise with reference tracking capability and robustness as increasing the low-frequency gain comes at the expense of a decrease in the magnitude slope and thus in the phase

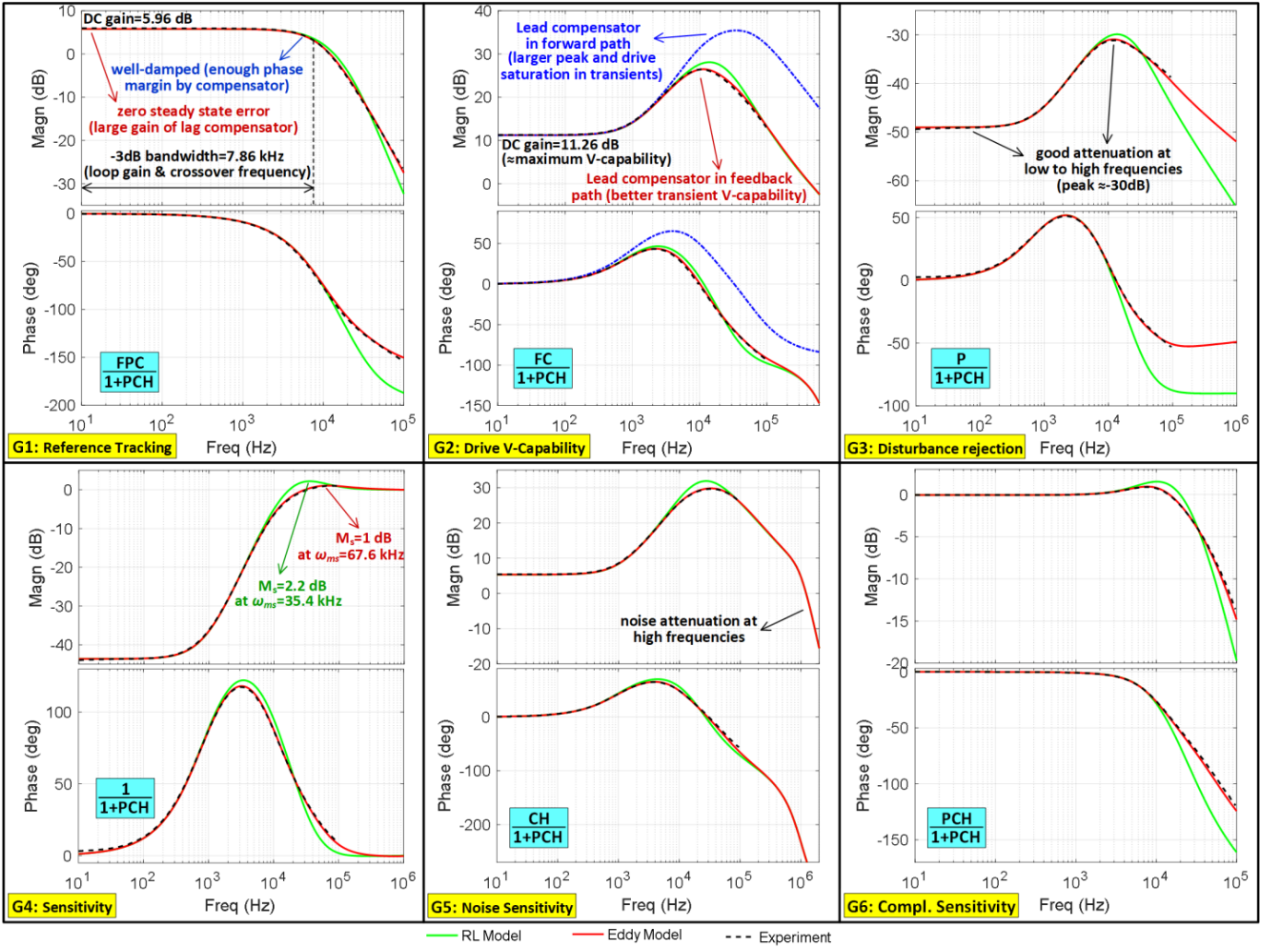

Fig. 6. Frequency response of the six gangs.

— RL Model  — Eddy Model  - - - Experiment

margin of the loop transmission around the crossover frequency. In other words, pushing down the output response to the disturbance (Gang 3) comes at the cost of an overshoot in the output response to the setpoint (Gang 1). The disturbances are effectively attenuated at low to high frequencies, as it can be observed in Fig. 6 that the magnitude peak is around -30 dB.

To obtain time responses of Gangs 3 to 6, extra equipment is not required to inject $D$ and $N$ signals to the specified locations. As the input impedance of the power op-amp is very large and the output impedance of the compensator op-amp is very low, according to the circuit shown in Fig. 7(c), approximated responses of Gangs 3 and 4 can be obtained by injecting the input signal to the non-inverting input of the power op-amp through a 10 kΩ resistor. If the inverse gains of the voltage divider ($v_{in}$ to $v_+$) and power op-amp ($v_+$ to $v_c$) are applied to the responses, $i_c$ and $v_c$ give the approximate responses for $G3=P/(1+PCH)$ and $G4=1/(1+PCH)$, respectively. The inverse of the total gain from $v_{in}$ to $v_c$ is 0.2, so if the magnitude of injected signal $v_{in}$ is 0.2 volt, the signals $i_c$ and $v_c$ give the unit step responses of Gangs 3 and 4. It is seen that the unit step response to the disturbance is effectively suppressed to

6 mv.

### D. Gang 4: Sensitivity

The sensitivity $S$ is the transfer function from the noise $N$ to the output $y$, or reference $R$ to the error for $F=1$

$$S = \frac{Y}{N} = \frac{1}{1+PCH} \qquad (21)$$

Typically, $S$ is zero at low frequencies, has a peak $M_s$ at a mid-frequency $\omega_{ms}$, and converges to unity at high frequencies. Sensitivity is a measure of the robustness of the control system to the variations of the parameters of the plant $P=H_e$ as the impact of variations of $T$ to $P$ is proportional to sensitivity $S$ as

$$\frac{dT}{dP} = S\frac{T}{P} \;\Rightarrow\; \frac{dT}{T} = S\frac{dP}{P} \qquad (22)$$

In our case, a very good robustness to the changes of coil resistance and inductance is an example, where the coil resistance is increases by 33% from 25 to 100 degrees Celsius while the inductance has minor changes up to 5%. If the sensitivity curve is harshly pushed down at low frequencies to obtain a smaller steady-state error and robust disturbance rejection, it pops up at mid frequencies resulting in a larger $M_s$; it is called waterbed effect and needs a trade-off. It is also reflected in the fact that $S+T=1$ if $F=H=1$. Usually, a value of $M_s$ smaller than 2dB or 3dB shows a satisfying design. Thanks to the sufficient phase margin of the loop, $M_s=1$ is obtained, as shown in Fig. 6. It can also be seen that if the RL model without eddy current dynamic was used, the value of $M_s$ had a significant discrepancy which can be very misleading in the design trade-offs. According to Fig. 7, the unit step response to the noise signal is effectively suppressed to 10 mv. It is also an indication of satisfying steady-state error elimination.

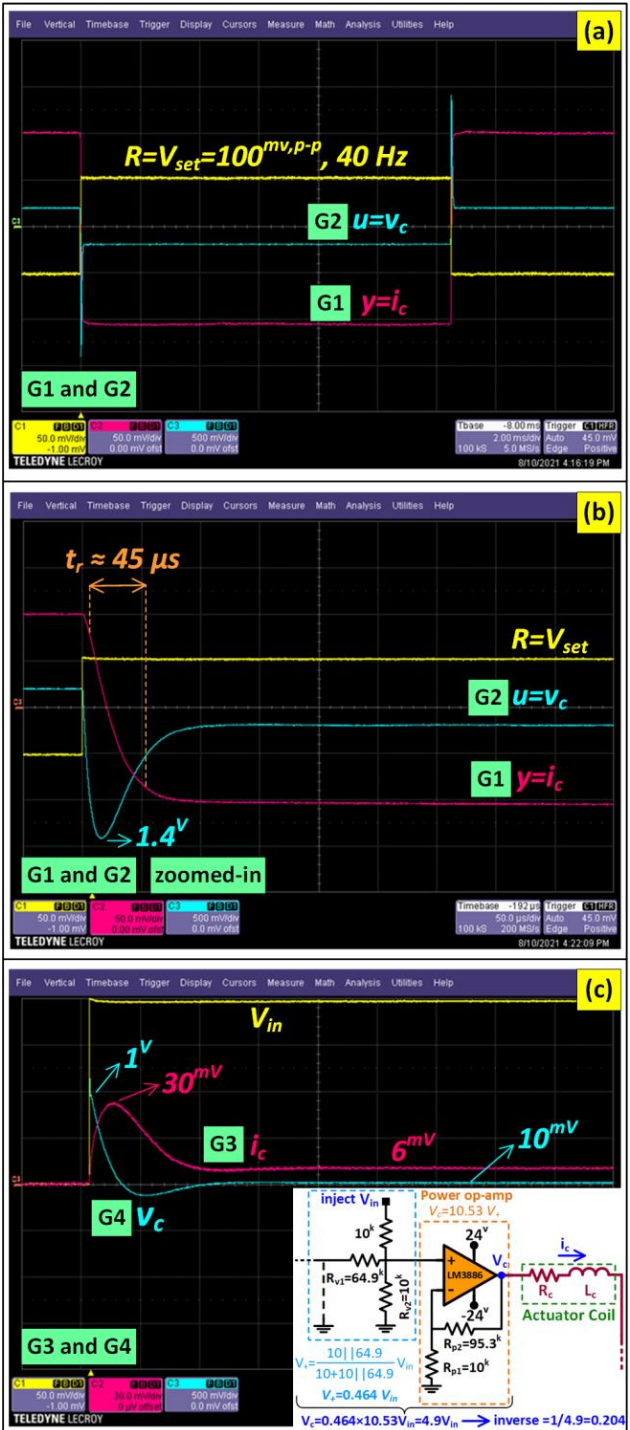

Fig. 7. Step responses of (a) Gang 1 and Gang 2, (b) Gang 1 and Gang 2 zoomed-in, and (c) Gang 3 and Gang 4.

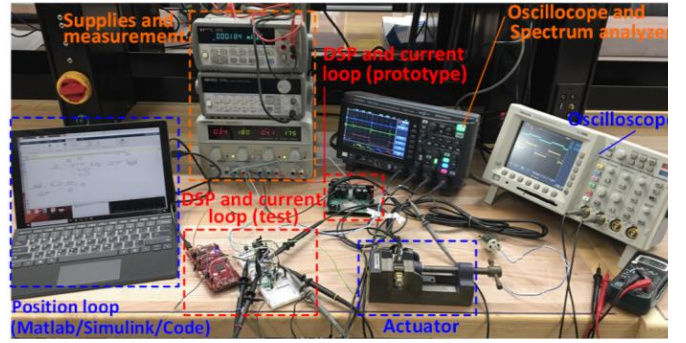

Fig. 8. Control setup

### E. Gang 5: Noise Sensitivity

The noise sensitivity is the transfer function from the noise $N$ to the drive output $U$.

$$S_N = \frac{U}{N} = \frac{CH}{1+PCH} \quad (23)$$

The system should be designed such that noise sensitivity is as small as possible so that the measurement noise is not amplified by the power op-amp, causing loss and drive saturation. As $S_N=CH\times S$, at high frequencies $S=1$ and so $S_N=CH$; thus, the pole-zero ratio of the lead compensator $\alpha$ should not be very large to avoid noise amplification. As shown in Fig. 6, a sufficient noise attenuation is obtained at high frequencies by a value of $\alpha=10$.

### F. Gang 6: Complementary Sensitivity

Complementary sensitivity is the transfer function from the disturbance $D$ to the drive output $U$.

$$S_{cm} = \frac{U}{D} = \frac{PCH}{1+PCH} \quad (24)$$

If $F=H=1$, $S_{cm}=T$. As $S+S_{cm}=1$, there is a compromise between $S$ and $S_{cm}$. It is shown in Fig. 6.

## IV. POLE PLACEMENT CONTROL USING VOLTAGE DRIVE

The position control can be performed with or without employing a current loop as the inner loop. In this section, the position loop is digitally implemented in a DSP. The Zero-Order-Hold (ZOH) sampling is performed at the frequency of $f_s$ up to 160 kHz. Bipolar ADCs with 16 bits of resolution are employed. If unipolar ADCs are used, it is required to deal with an offset by an extra op-amp circuit. The position sensor returns a voltage as a function of position, and its inverse function is implemented in the DSP. As the bandwidth of the position loop should be around or not much larger than the bandwidth of the actuator to avoid drive saturation, pole placement position control is employed for desired poles having a natural frequency of $\omega_n=2\pi500 \ rad$. The experimental control setup is shown in Fig. 8. As shown in Fig. 9, the pole placement control is performed using the power op-amp as a voltage drive. To effectively use the resolution of the DAC, a voltage divider with a gain of 0.4 is used such that $\pm5v$ at the DAC translates to $\pm21v$ at the output of power op-amp ($\pm5\times0.4\times10.53=\pm21$). The coil voltage is measured by ADC through a voltage divider. Also, the current can be measured using the output of the buffer sent to an ADC, or it can be estimated by a state observer. The circuits gains are canceled out in the DSP by their inverse values so that the physical model of the actuator can be used for control system design without requiring any gain modification.

### A. Employed Model

By ignoring the fractional-order dynamic of eddy currents, an integer-order linearized model, whose block diagram is shown in the Appendix, is obtained to be used in the pole placement control as follows

$$\begin{cases} v_c = k_t \omega_r + L_{c0} \dfrac{di_c}{dt} + R\, i_c \\ J \dfrac{d^2\theta}{dt^2} + K_d \dfrac{d\theta}{dt} + K_s \theta = k_t\, i_c \end{cases} \quad (25)$$

It can be represented as a third-order state-space model as

$$\begin{bmatrix} \dot{\theta} \\ \dot{\omega}_r \\ \dot{i}_c \end{bmatrix} = \begin{bmatrix} 0 & 1 & 0 \\ -\dfrac{K_s}{J} & -\dfrac{K_d}{J} & \dfrac{k_t}{J} \\ 0 & -\dfrac{k_t}{L_{c0}} & -\dfrac{R}{L_{c0}} \end{bmatrix} \begin{bmatrix} \theta \\ \omega_r \\ i_c \end{bmatrix} + \begin{bmatrix} 0 \\ 0 \\ \dfrac{1}{L_{c0}} \end{bmatrix} v_c \quad (26)$$

### B. Full-State Feedback Control in Time Domain

As shown in Fig. 9, full-state feedback is obtained by substituting $u=r-K\,\delta x$ and $r=G\,\theta_{ref}$ as

$$\frac{d}{dt}\delta x(t) = (A-BK)\,\delta x(t) + B\, r(t) \quad (27)$$

The eigenvalues of matrix $A_{cl}=A-BK$ determine the closed-loop dynamic. The gain vector $K=[k_1, k_2, k_n]$ is obtained by pole-placement using Ackermann's formula as

$$\begin{cases} K = \begin{bmatrix} 0 & 0 & 1 \end{bmatrix}_{1\times 3} M_c^{-1}\, \phi_d\left(A_{3\times 3}\right) \\ M_c = \begin{bmatrix} B & AB & A^2B \end{bmatrix} \end{cases} \quad (28)$$

where $M_c$ is the controllability matrix and $\varphi_d$ is the desired characteristic polynomial whose roots are the desired eigenvalues $\lambda_1$, $\lambda_2$ and $\lambda_3$ of closed-loop dynamic $A_{cl}=A-BK$ which are chosen to be on a circle with a radius of $\omega_n = 2\pi f_n$ and with damping of $\zeta$ as $-\omega_n$ and $-\xi\omega_n \pm j\omega_n\sqrt{1-\xi^2}$. It leads to the following desired characteristic polynomial

$$\begin{cases} \varphi_d(\lambda) = (\lambda - \lambda_1)(\lambda - \lambda_2)(\lambda - \lambda_3) \\ \varphi_d(\lambda) = (\lambda^2 + 2\zeta\omega_n\lambda + \omega_n^2)(\lambda + \omega_n) \end{cases} \quad (29)$$

The input gain for position tracking ($C=[1\ 0\ 0]^t$) is obtained as

$$G = -[C(A-BK)^{-1}B]^{-1} \quad (30)$$

### C. Full-Order State Estimator

Position and current can be directly measured or estimated, and velocity is estimated. When there are noise problems and unmodeled dynamics, as in our case where eddy current dynamics are ignored, a full-order observer might be preferred over a reduced-order one. The estimator dynamic is as

$$\frac{d}{dt}\delta\hat{x}(t) = A\,\delta\hat{x}(t) + Bu(t) + L(y(t)-\hat{y}(t)) \quad (31)$$

$$\hat{y}(t) = C\,\delta\hat{x}(t) \quad (32)$$

Substituting for $\hat{y}$ in (15) results in

$$\frac{d}{dt}\delta\hat{x}(t) = (A-LC)\,\delta\hat{x}(t) + \begin{bmatrix} B & L \end{bmatrix}\begin{bmatrix} u(t) \\ y(t) \end{bmatrix} \quad (33)$$

where $A_e=A-LC$ forms the closed-loop dynamics of the estimator. The pole placement can be done for the estimator using Ackermann's formula to obtain the gain vector $L=[L_1\ L_2\ L_3]^t$ as

$$\begin{cases} L = \phi_e(A_{3\times3})M_o^{-1}\begin{bmatrix} 0 & 0 & 1 \end{bmatrix}_{1\times3}^t \\ M_o = \begin{bmatrix} C \\ CA \\ CA^2 \end{bmatrix} \end{cases} \quad (34)$$

where $M_o$ is the observability matrix, and $\varphi_e(\lambda)$ is the desired characteristic polynomial whose roots are the desired eigenvalues of estimator dynamic $A_e=A-LC$ which are chosen to be around 5 to 10 times faster than the controller. For example, locating them at $-10\omega_n$, can be a good choice as it is still within the bandwidth of the sensors. It leads to the following characteristic polynomial

$$\varphi_e(\lambda) = (\lambda + 10\omega_n)^3 \quad (35)$$

Using the Forward Euler method, by substituting $d/dt$ with $(z-1)/T_s$, the Z-transform and the discrete-time equation of the estimator is obtained as

$$\hat{x}(k) = (I + T_s A_e)\hat{x}(k-1) + T_s B v_c(k-1) + T_s L\, y(k-1) \quad (36)$$

where $T_s=1/f_s$ is the sampling time. It can be easily implemented into the DSP. Another state estimation technique is to employ a full-order observer where only the unmeasured states (velocity) are taken from the observer, and the measured states (position and current) are directly taken from the sensor. In this method, model uncertainties can be more efficiently suppressed in velocity estimations.

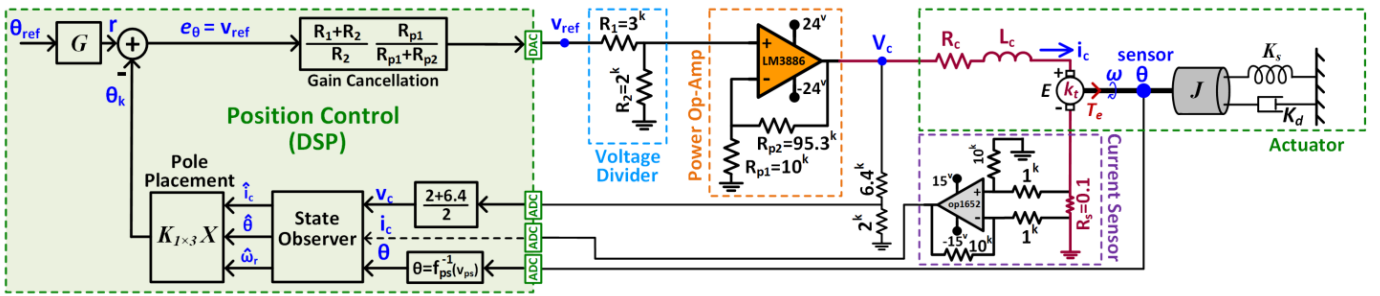

Fig. 9. Pole placement with voltage drive: (a) block diagram of the control system, and (b) block diagram of the linearized electromechanical model.

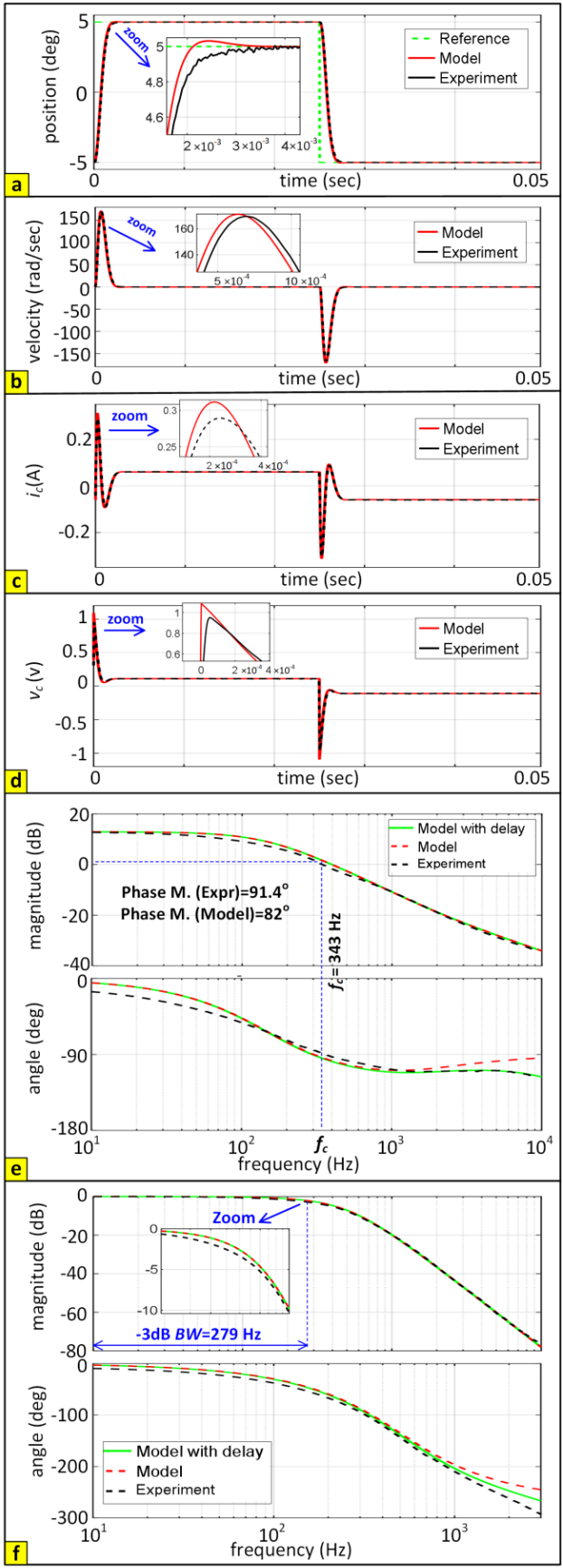

Fig. 10. Pole placement with voltage drive: (a)-(d) step responses of position, velocity, current and voltage, and (e)-(f) bode of loop transmission and closed-loop.

## D. Design of the Compensator

The compensator is the combined controller and estimator with input $y(t)$ and outputs $u(t)$. If $r=0$, the dynamics is obtained by substituting $u = -K\,\delta\hat{x}$ and $\hat{y} = C\,\delta\hat{x}$ in (15) as

$$\begin{cases} \dfrac{d}{dt}\delta\hat{x}(t) = (A - BK - LC)\delta\hat{x}(t) + L\,y(t) \\ u = -K\delta\hat{x}(t) \end{cases} \quad (37)$$

where $A_c = A - BK - LC$, $B_c = L$ and $C_c = -K$. Its dynamics are obtained as eigenvalues of $A_c = A - BK - LC$, which need to be checked for stability. The closed-loop dynamic is as

$$\frac{d}{dt}\begin{bmatrix} \delta x(t) \\ \delta \hat{x}(t) \end{bmatrix} = \begin{bmatrix} A & -BK \\ LC & A - BK - LC \end{bmatrix} \begin{bmatrix} \delta x(t) \\ \delta \hat{x}(t) \end{bmatrix} \quad (38)$$

The characteristic polynomial of the compensator is $|\lambda I-(A-BK)|\times|\lambda I-(A-LC)|=0$, and so the six eigenvalues of the above system are the same as the three eigenvalues of $A_{cl}=A-BK$ and the three eigenvalues of $A_e=A-LC$ taken together. This fact is called the separation principle that enables us with the independent design of controller and estimator.

## E. Design, Simulation, and Experiment

The plant is controllable and observable as $M_c$ and $M_o$ are full rank matrices. The eigenvalues of $A_{cl}$ are chosen by $\omega_n=2\pi f_n=2\pi500$ rad/sec and damping of $\zeta=0.8$. The feedback and the input gains are obtained as $K=[5.3636, 0.0031, 0.3437]$ and $G=6.8664$. The eigenvalues of estimator dynamic $A_e$ are chosen to be 5 to 10 times faster than controller. For a response that is 10 times faster, the value of estimator gain $L$ is obtained as [8.73e4, 2.34e9, 1.15e7]. Also, the compensator is stable as the eigenvalues of $A-BK-LC$ are -42069 and -26703±11066i.

Figs. 10(a)-(d) shows both simulation and experimental results for a square wave reference with a magnitude of ±5 degrees and a frequency of 20 $Hz$. The steady-state error is almost zero, voltage and current are within limits, and the experimental results are close to those expected from simulations. As shown in Fig. 10(a), a small discrepancy is observed in the reference tracking results; the simulation predicts a small overshoot which is expected from the desired damping, while the experiment does not illustrate any overshoot. It can also be explained by the closed-loop frequency response given in Fig. 10(f) that the experimental result shows a more damped system. This discrepancy can be explained by non-modeled dynamics such as friction as well as eddy-currents; as shown in Fig. 10(e), the phase margin of the real system is a bit larger than the model, i.e., a smaller overshoot. Although the obtained phase margin looks good, the closed-loop response is a function of temperature-dependent elements like coil resistance. The closed-loop transfer function ($r$ to $\theta$) is obtained as

$$T = GC(SI - [A - BK])^{-1}B \quad (39)$$

Also, the loop transmission, which is the transfer function from $u$ to the comparison point $\theta_k$ is obtained as

*A.  Employed Model*

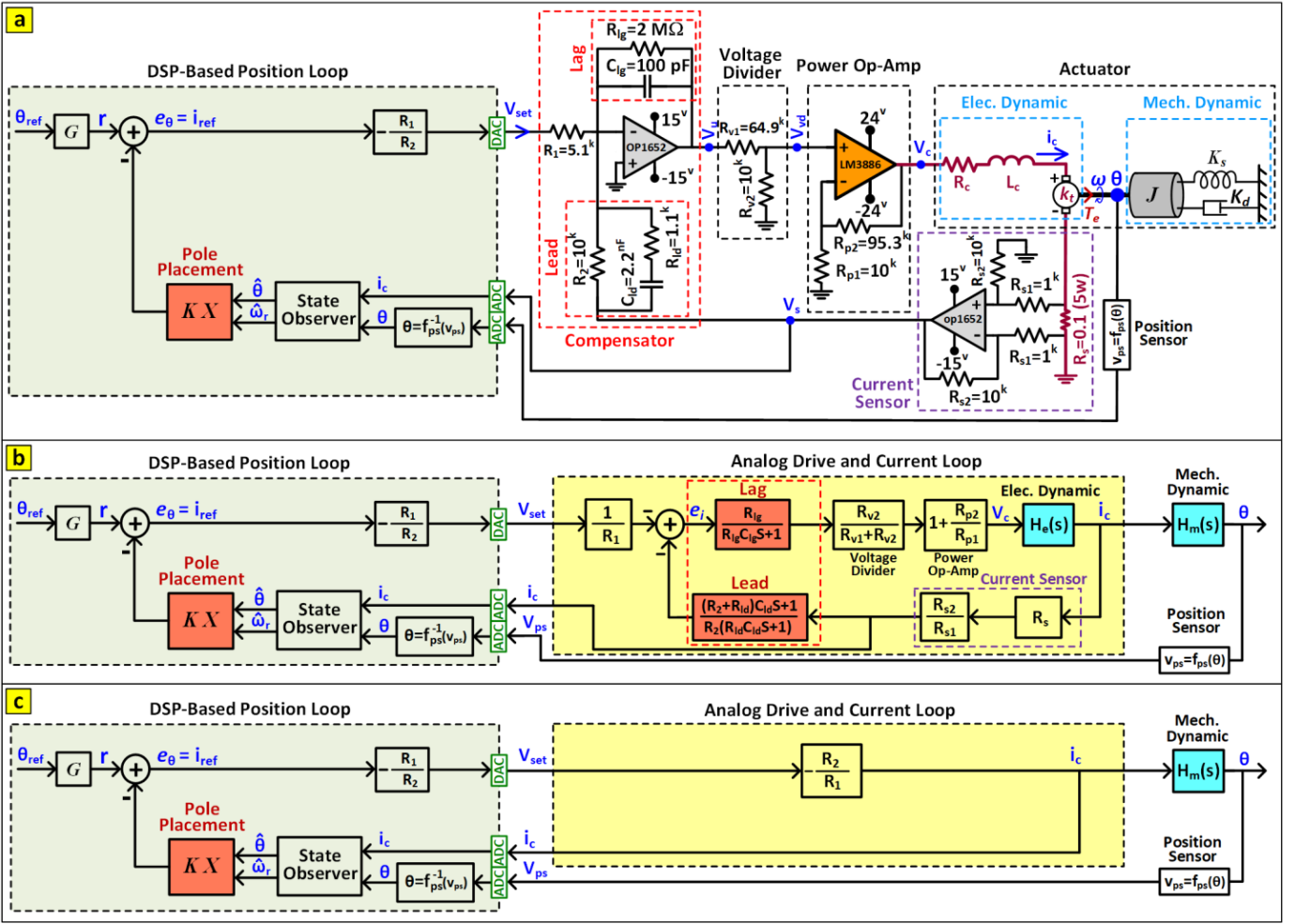

Fig. 11. Pole placement with current drive: (a) implemented control system, (b) block diagram of the current and position loops, and (c) treating the high bandwidth current loop as a DC gain.

$$L = \frac{\theta_k}{I_{ref}} = K(SI - A)^{-1}B \qquad (40)$$

Corrections of the delays due to ADC and computation time can be performed by the term $e^{-sT_d}$ where $T_d$ is the delay.

## V.  POLE PLACEMENT CONTROL USING CURRENT DRIVE

As shown in Fig. 11(a), using a high-bandwidth current loop, the electrical dynamic of the actuator, including its time constant and complicated dynamics such as eddy currents, can be eliminated, leaving a faster plant having less complexities. Then, the current or torque can instantaneously be commanded by the position loop. The bandwidth of the current loop is around 7.86 kHz, while the desired bandwidth of the position loop is less than 500 Hz. Therefore, the block diagram of the control system shown in Fig. 11(b) can be simplified to the one shown in Fig. 11(c) in which the current loop is seen as its DC gain from the position loop. This gain is canceled out by its inverse in the DSP in order to employ the original plant model in the design process.

Eliminating the electrical dynamic, the model (10) is reduced to the second-order mechanical dynamic as

$$\begin{bmatrix} \dot{\theta} \\ \dot{\omega}_r \end{bmatrix} = \begin{bmatrix} 0 & 1 \\ -\dfrac{K_s}{J} & -\dfrac{K_d}{J} \end{bmatrix} \begin{bmatrix} \theta \\ \omega_r \end{bmatrix} + \begin{bmatrix} 0 \\ \dfrac{k_t}{J} \end{bmatrix} i_c \qquad (41)$$

### B.  Full-State Feedback Control in Time Domain

Exactly like the previous section, the feedback gains $K=[k_1 \ k_2]$ and the unitary input gain $G$ are obtained where $C=[1 \ 0]^t$.

### C.  Reduced-Order Estimator

The available states do not need to be estimated by the observer. Reduced-order observers are computationally more efficient, may converge faster, and have higher bandwidth. For the current drive, a reduced-order observer is employed to estimate velocity. The model can be partitioned based on the measured states $X_1=\theta$ and unmeasured ones $X_2=\omega_r$ as

$$\begin{cases} \begin{bmatrix} \dot{X}_1 \\ \dot{X}_2 \end{bmatrix} = \begin{bmatrix} A_{11} & A_{12} \\ A_{21} & A_{22} \end{bmatrix} \begin{bmatrix} X_1 \\ X_2 \end{bmatrix} + \begin{bmatrix} B_1 \\ B_2 \end{bmatrix} i_c \\ y = [I \ 0] \begin{bmatrix} X_1 \\ X_2 \end{bmatrix} \end{cases} \quad (42)$$

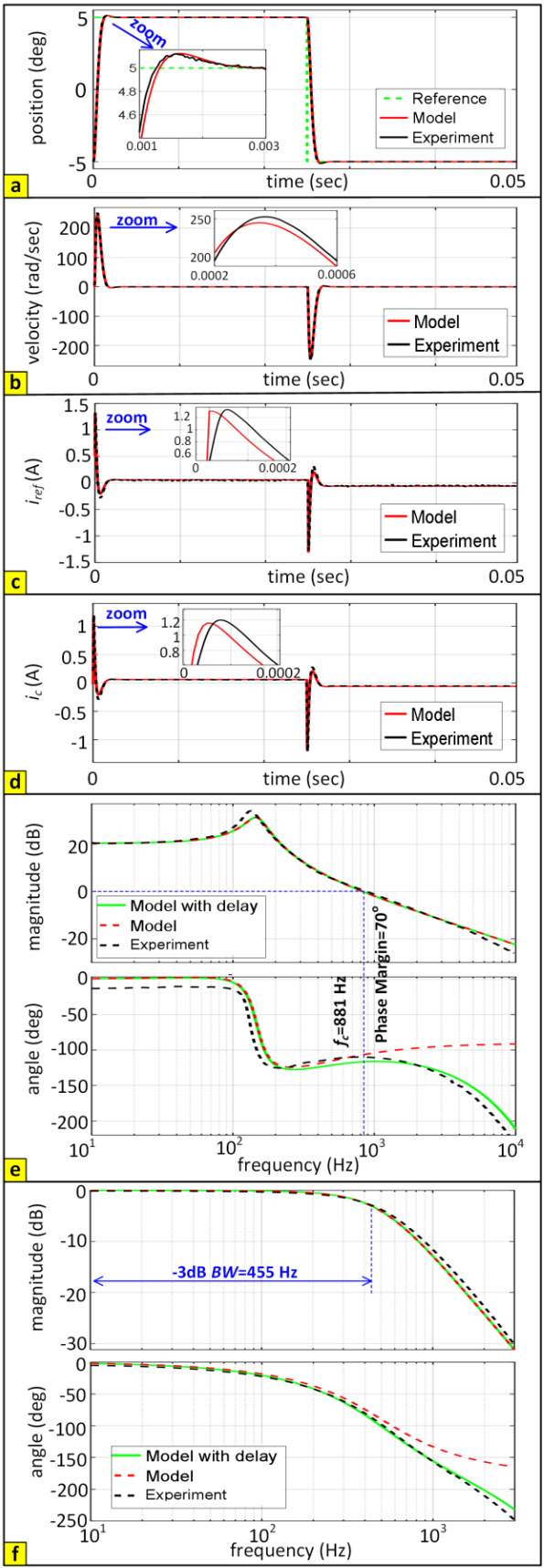

Fig. 12. Pole placement with current drive: (a)-(d) step responses of position, velocity, current command, and coil current. Also, frequency response of (e) loop transmission, and (f) the closed-loop system

The estimator in terms of the new state z can be expressed as

$$\begin{cases} \dot{Z} = \hat{A} Z + \hat{B} y + \hat{F} i_c \\ X_2 = Z + L y \end{cases} \quad (43)$$

whose parameters are obtained as

$$\begin{cases} \hat{A} = A_{22} - L A_{12} \\ \hat{B} = \hat{A} L + A_{21} - L A_{11} \\ \hat{F} = B_2 - L B_1 \end{cases} \quad (44)$$

Thus, the characteristic polynomial of $\hat{A}=-k_d/J-L$ is obtained whose characteristic polynomial is $\varphi(\lambda)=|\lambda I-\hat{A}|=\lambda+k_d/J+L$. Also, the bandwidth of the estimator is $\lambda_0$, so the desired pole is $-\lambda_0$ and the desired characteristic polynomial is $\varphi_e(\lambda)=\lambda+\lambda_0$. Thus, the estimator gain is obtained as $L=\lambda_0-k_d/J$. Also, Ackermann's formula (34) can be used to obtain estimator's gain by substituting $A$ with $A_{22}$ and $C$ with $A_{12}$ as

$$\begin{cases} L = \phi_e(A_{22}) M_o^{-1} [1]^t \\ M_o = [A_{12}] \end{cases} \quad (45)$$

It gets to the same value for $L$. Substituting $L$ in (40) leads to

$$\begin{cases} \hat{A} = -\lambda_0 \\ \hat{B} = -\left( \lambda_0^2 - \dfrac{k_d \lambda_0}{J} + \dfrac{k_x}{J} \right) \\ \hat{F} = \dfrac{k_t}{J} \end{cases} \quad (46)$$

Finally, the velocity is obtained as $\omega_r=z+L\theta$. Using the Forward Euler, the discrete-time equations are obtained for DSP implementation as

$$\begin{cases} Z(k) = (I + T_s \hat{A}) Z(k-1) + T_s \hat{B} \theta(k-1) + T_s \hat{F} i_c(k-1) \\ \omega_r(k) = Z(k) + L \theta(k) \end{cases} \quad (47)$$

The estimator bandwidth is set to $\lambda_0=10\omega_n$.

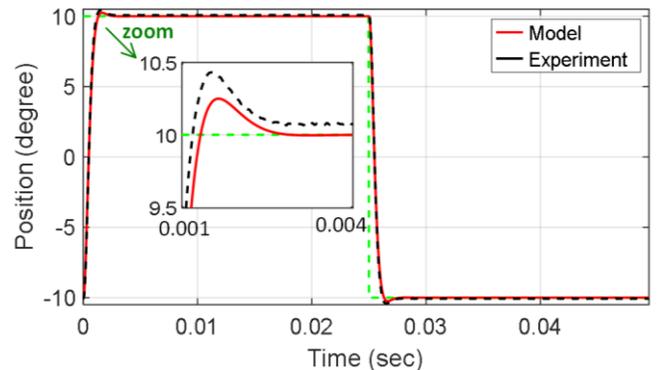

Fig. 13. Large-signal response of the pole placement with current drive.

## D. Compensator

It can be shown that the characteristic equation of the compensator $|\lambda I - (A - BK)| \times |\lambda I - (A_{22} - LA_{12})| = 0$, so the controller dynamic $A_{cl} = A - BK$ and the estimator dynamic $\hat{A} = A_{22} - LA_{12}$ can be designed independently.

## E. Design, Simulation, and Experiment

The desired closed-loop poles $\lambda_1$ and $\lambda_2$ are chosen to have a natural frequency of $\omega_n = 2\pi f_n = 1000\pi$ rad/sec, and damping of $\zeta = 0.8$ as $-\xi\omega_n \pm j\omega_n\sqrt{1-\xi^2}$, so the desired characteristics polynomial is $\varphi_d(\lambda) = (\lambda - \lambda_1)(\lambda - \lambda_2) = \lambda^2 + 2\zeta\omega_n\lambda + \omega_n^2$. The feedback gains and the unitary gain are obtained as $K = [7.124, 0.0037]$ and $G = 7.806$. Then, the estimator gain $L$ is obtained as 31118.

The step responses of position, velocity, current command (scaled DAC output), and coil current are shown in Figs. 12(a)-(d). The reference tracking and the performance of the current loop are very good. Not only are the results as expected from the experiment, but also, they correlate well with the simulations from the model. It can be observed that, compared to voltage drive control, the control system design using the current drive is more accurate, which is due to the elimination of electrical dynamics, including eddy currents and back-emf impact. Also, the elimination of the temperature-dependent resistance of the coil adds to the robustness of the system. As shown in Fig. 12(e) and (f), the performance of the system is checked in the frequency domain, illustrating a sufficient phase margin of 70 degrees and a -3dB bandwidth of 455 Hz, which is higher than the bandwidth of the pole placement control with voltage drive. As shown in Fig. 13, there is a steady error and a bit larger overshoot in the large-signal reference tracking result of the control system for a reference amplitude of 10 degrees.

Although current loop dynamics can be reduced to a simple gain for controller design, it can be included in the model to gain a higher accuracy in the designs and simulations. If the transfer function $H_{CL}$ is the closed-loop response of the current loop (Gang 1) multiplied by the inverse of its DC gain to have a unity DC gain on total, the control effort, instead of being $U = G\theta_{ref} - K \delta X$, will be $U = H_{CL}(R - K \delta X)$. The transfer function of the plant from $u$ to the states as outputs is the 1-input 2-output system $G_m = C(SI - A)^{-1}B$ where $C = I_{2 \times 2}$. The difference between $G_m$ and $H_m$ is that $G_m$ is a 2-by-1 matrix that outputs both position and velocity. Thus, a closed-loop system incorporating the current-loop dynamics is obtained as

$$X = G_m \overbrace{H_{CL}(G\theta_{ref} - KX)}^{u} \Rightarrow$$
$$\frac{X}{\theta_{ref}} = (I_{2 \times 2} + G_m H_{CL}K)^{-1}G_m H_{CL}G \tag{48}$$

## VI. NONLINEAR CONTROL BY FEEDBACK LINEARIZATION

The linear control system works well for small-signal setpoints while, for large-signal maneuvers, they can result in unwanted inaccuracies like steady-state error, large overshoots, and even instability in severe cases. Feedback linearization shown in Fig. 14(a) can be powerful in eliminating the nonlinearities of the system, yet it requires measuring or estimating the state variable as well as a very accurate model of the plant that we developed. The current loop is employed to get a faster response and to get rid of the complexities and fractional-order elements of the electrical dynamics. Then, we only deal with the nonlinear model of the mechanical dynamics as shown in the Appendix. As the the nonlinear characteristics of restoration torque and the electromagnetic torque are functions of position, by substituting $\theta = \beta - \pi/2$, the nonlinear electromechanical model is obtained as

$$\begin{cases} Elec: \ v_c = k_b \omega_r \cos\theta + R_c \ i_c + L_c \dfrac{di_c}{dt} \\ Mech: J \dfrac{d^2\theta}{dt^2} + K_d \dfrac{d\theta}{dt} + k_{rest} \sin 2\theta = k_t \ i_c \cos\theta \end{cases} \tag{49}$$

## A. Feedback Linearization

Feedback linearization can be implemented for a plant if its state-space model can be written in the companion form as

$$\begin{cases} \dot{x}_1 = x_2 \\ \dot{x}_2 = x_3 \\ ... \\ \dot{x}_n = f(x_1,...,x_n) + g(x_1,...,x_n)u(t) = v(t) \end{cases} \tag{50}$$

where $f(x)$ and $g(x)$ are nonlinear functions of the states and $u(t) = i_c(t)$ is the input. All of the states need to be measured or estimated in order to calculate $f$ and $g$. Then, the following nonlinear transformation is used to cancel out the nonlinearities.

$$u(t) = \frac{1}{g(x_1,...,x_n)}\big[v(t) - f(x_1,...,x_n)\big] \tag{51}$$

It results in a linear system with $n$ poles at the origin and with the new input $v(t)$, to which linear control techniques can be applied. The nonlinear mechanical dynamic can be written as

$$\begin{cases} \dot{\theta} = \omega_r \\ \dot{\omega}_r = -\dfrac{k_d \omega_r + k_{rest}\sin 2\theta}{J} + \dfrac{k_t \cos\theta}{J}i_c = f + g \ i_c = v \end{cases} \tag{52}$$

where functions $f$ and $g$ are obtained as

$$\begin{cases} f(\theta, \omega_r) = -\dfrac{k_d \omega_r + k_{rest}\sin 2\theta}{J} \\ g(\theta, \omega_r) = \dfrac{k_t \cos\theta}{J} \end{cases} \tag{53}$$

The nonlinear transformation at the input is as

$$i_c(t) = \frac{1}{g(\theta, \omega_r)}[v(t) - f(\theta, \omega_r)] \tag{54}$$

The linear system and its state-space form are obtained as

$$\ddot{\theta} = v \ \Rightarrow \ H'_m = \frac{\theta(s)}{v(s)} = \frac{1}{s^2} \tag{55}$$

$$\begin{bmatrix} \dot{\theta} \\ \dot{\omega}_r \end{bmatrix} = \begin{bmatrix} 0 & 1 \\ 0 & 0 \end{bmatrix}\begin{bmatrix} \theta \\ \omega_r \end{bmatrix} + \begin{bmatrix} 0 \\ 1 \end{bmatrix}v \tag{56}$$

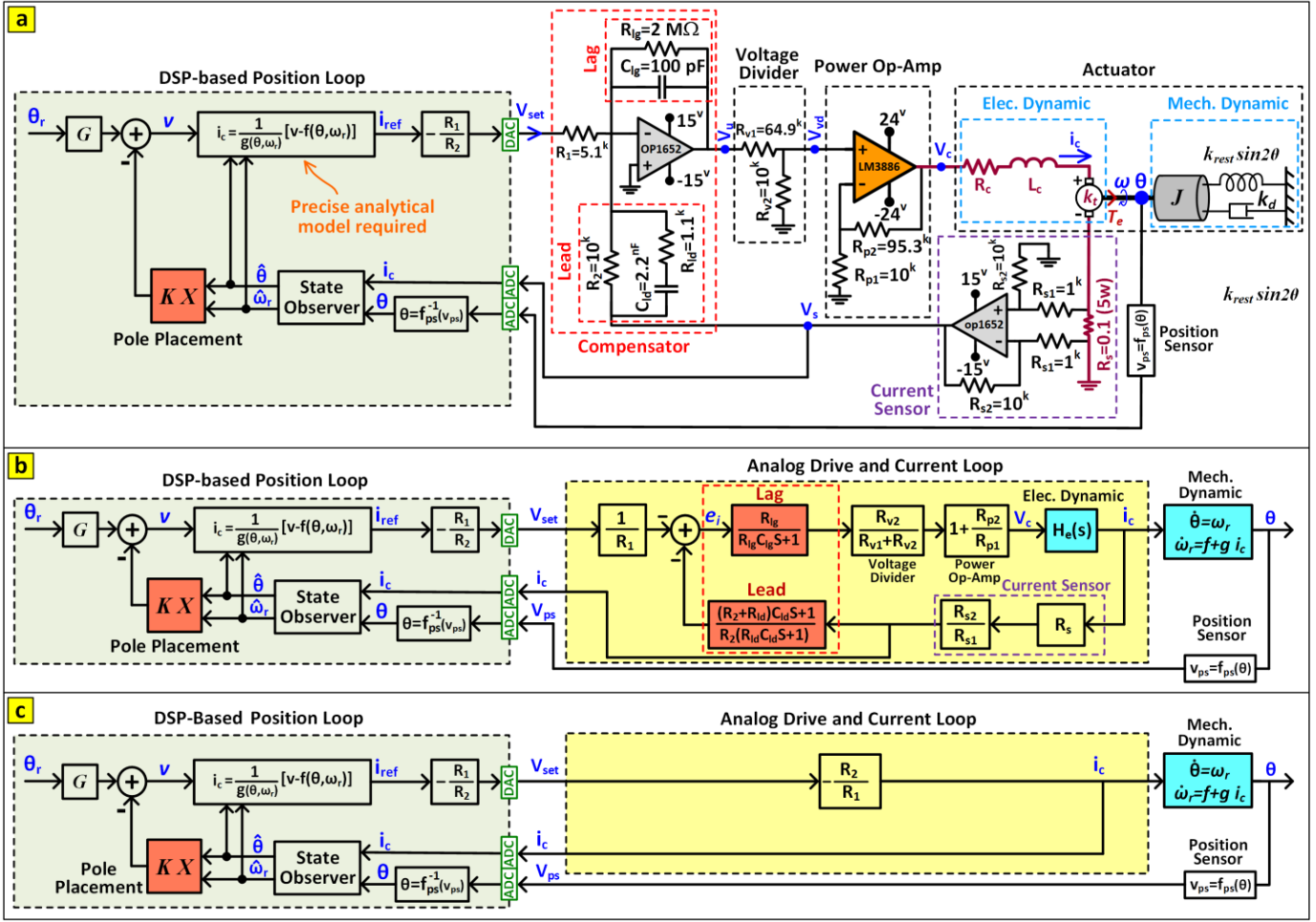

Fig. 14. Nonlinear control: (a) implemented control system, (b) block diagram of the feedback linearization control, and (c) treating the high bandwidth current loop as a DC gain.

The block diagram of the feedback linearization is shown in Fig. 14(b). As shown in Fig. 14(c), the current loop is considered as its DC gain since its bandwidth is much larger than the bandwidth of the position loop. However, its dynamic is included in the simulations.

## B. Pole Placement in Time Domain

The desired closed-loop poles $\lambda_1$ and $\lambda_2$ are chosen to have a natural frequency of $\omega_n = 2\pi f_n$ and damping of $\zeta$ as $-\xi\omega_n \pm j\omega_n\sqrt{1-\xi^2}$, so the desired characteristics polynomial is $\varphi_d(\lambda) = \lambda^2 + 2\xi\omega_n\lambda + \omega_n^2$. The matrices $A$, $B$, and $C$ are obtained as

$$A = \begin{bmatrix} 0 & 1 \\ 0 & 0 \end{bmatrix}, \quad B = \begin{bmatrix} 0 \\ 1 \end{bmatrix}, \quad C = [1 \ 0] \tag{57}$$

The feedback gains $K=[k_1 \ k_2]$ for position and velocity obtained by Ackermann's formula as well, as the unitary input gain is obtained as

$$\begin{cases} k_1 = \lambda_1\lambda_2 = \omega_n^2 \\ k_2 = -(\lambda_1 + \lambda_2) = 2\xi\omega_n \\ G = \omega_n^2 \end{cases} \tag{58}$$

The velocity observation is performed using a derivate plus a low-pass filter which is kind of like the reduced-order observer used in the pole placement with the current drive. Also, as shown in Fig. 15, it can mathematically be shown that the transfer function of the loop transmission is almost a double integrator (linearized system from $v$ to $\theta$) in series with a PD compensator in the feedback loop as

$$L \approx \frac{\theta_k}{v} = \frac{\omega_n^2 + 2\xi\omega_n s}{s^2} \tag{59}$$

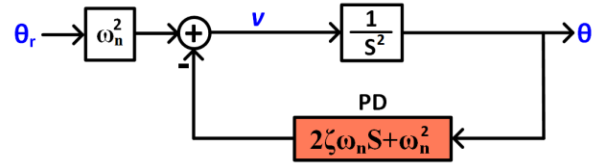

Fig. 15. Simplified block diagram of the nonlinear control system.

Therefore, the closed-loop system is obtained as

$$\frac{\theta}{\theta_{ref}} = \frac{G/s^2}{1+L} = \frac{\omega_n^2}{s^2 + 2\xi\omega_n s + \omega_n^2} \tag{60}$$

## C. Design, Simulation, and Experiment

The desired closed-loop poles have a natural frequency of $\omega_n = 2\pi f_n = 1000\pi$ rad/sec and damping of $\zeta = 0.8$. The step responses of position, velocity, current command (scaled

DAC output), and coil current for a large-signal command with an amplitude of 10 degrees are shown in Fig. 16. A comparison is also made with the simulations. Thanks to the accuracy of the developed nonlinear model, the nonlinear control technique works as expected and correlates well with the simulations. Contrary to the linear control system, the developed nonlinear control leaves a zero steady-state error in large signal tracking.

The system performance is also checked in the frequency domain given in Fig. 17. The frequency response of the system from the signal $v$ to the position is very close to a double integrator with a slope of -40 dB/dec as given in Fig. 17(a); it should be noted that its gain is attenuated for measurements by SR785 digital signal analyzer, and also a delay is observed in the phase which due to sampling and computations. It can also be seen in Fig. 17(b)-(c) that the loop transmission is almost a double integrator in series with a PD compensator. A sufficient phase margin of 59 degrees is obtained as well. As shown in Fig. 17(d) shows a bandwidth of 413 Hz, which is closed to the one obtained by the linear control system with the current drive. As shown in. Fig. 17(e), the maximum sensitivity of the control loop is $M_s$=2.4 dB, showing sufficient robustness.

## VII. CONCLUSION

In this two-part paper, a comprehensive exploration of modeling, identification, drive, and control system design for a rotary actuator with magnetic restoration is unveiled. In Part I, innovative nonlinear and linearized electromechanical models of the actuator are introduced that intricately account for eddy currents in both the laminations and the magnet, along with the effects of pre-sliding friction on system damping and stiffness. The proposed model surpasses existing models in terms of accuracy and simulation capabilities. The accuracy of the proposed model is verified by FEM and an experimental prototype. Subsequently, a systematic and effective identification procedure for extracting the parameters of electrical and mechanical dynamics is presented.

In part II, an op-amp-based analog drive circuit is proposed, modeled, designed, and employed as the current control loop. A detailed model of the drive circuit, along with simplified versions for application-specific accuracy requirements, are developed. Practical trade-offs and various aspects of the current control loop are thoroughly investigated and discussed. The developed models for both the actuator and the drive circuit play a significant role in enhancing the effectiveness of control system designs and the accuracy of the simulation platform. Through this, the evident superiority of the proposed model over existing approaches is demonstrated.

Subsequently, three DSP-based position control techniques are designed and implemented. Initially, a pole placement position control with voltage drive is developed, showcasing commendable performance in simple applications but revealing limitations in accuracy and robustness for more sophisticated control demands. Following this, leveraging the high-bandwidth current control loop, the intricacies and nonlinearities of the electrical dynamics are eliminated, leading to the implementation of a pole placement position control with current drive which demonstrates improved accuracy and robustness, albeit still lacking effectiveness for large-signal tracking. The high precision of the developed model makes it feasible to implement a feedback linearization control which is then employed for the nonlinear control of the actuator, particularly beneficial for applications requiring large signal tracking. The developed observers showed effectiveness in estimating the unmeasured states. The evaluation of control system designs encompasses key indices in both time response and frequency domain.

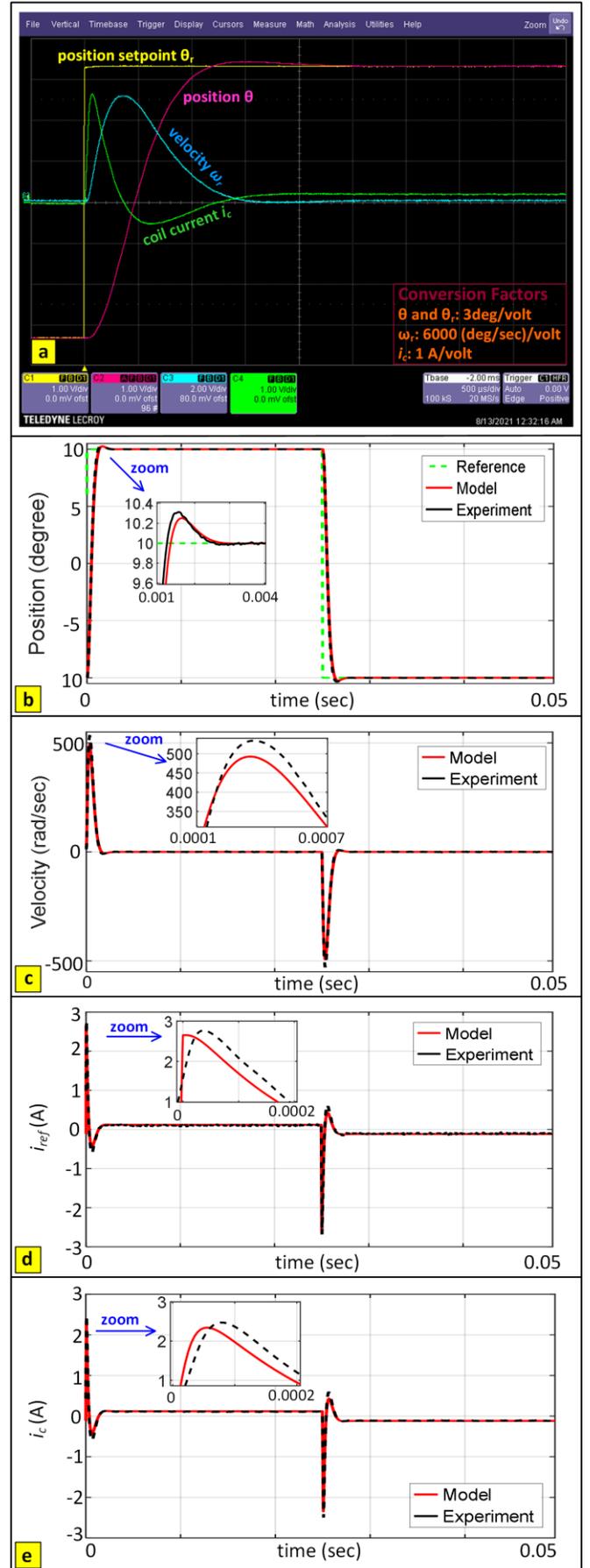

Fig. 16. Nonlinear control: (a) time responses, and (b)-(e) full-period waveforms and comparison with model for position, velocity, current command, and coil current.

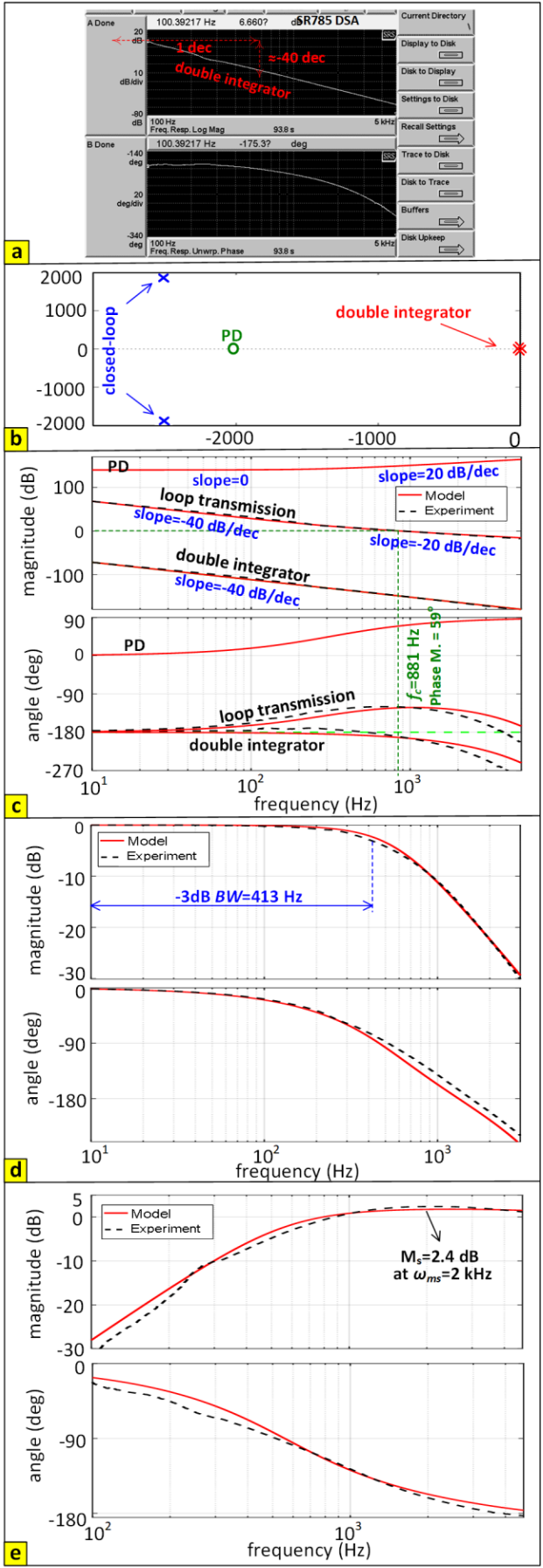

Fig. 17. Frequency domain of nonlinear control: (a) double integrator, (b) pole locations, (c) loop components, (d) closed loop, and (e) sensitivity.

|                   | Voltage Drive | Current Drive | Nonlinear |
|-------------------|:-------------:|:-------------:|:---------:|
| **Bandwidth**     | 2             | 1             | 1         |
| **Robustness**    | 2             | 1             | 1         |
| **Accuracy**      | 3             | 2             | 1         |
| **Small Signal**  | 2             | 1             | 1         |
| **Large Signal**  | 3             | 2             | 1         |
| **Simplicity/Cost** | 1           | 2             | 3         |

In Table I, a comparative analysis and ranking of the three position control systems are presented, considering various indices relevant to diverse applications. The position control system with voltage drive exhibits advantages in simplicity and cost-effectiveness, making it a viable choice for applications where extensive bandwidth or high accuracy is not a primary concern. On the other hand, both the position control with current drive and the nonlinear control deliver faster, more robust, and more accurate control systems. However, they necessitate current control drives, faster processors, and higher sampling rates, leading to increased costs. It is noteworthy that the implementation of nonlinear position control via feedback linearization requires the development of an exceptionally precise model of the system.

It is noteworthy that while complex models offer advantages in terms of accuracy, simulations, predictions, designs, and advanced controls, they do come with associated costs and drawbacks, such as the necessity for faster processors and higher sampling rates for analog to digital converters (ADCs). Designers must carefully weigh these factors against the specific requirements of the application, allowing them to choose between advanced models and more streamlined options, such as the simplified version of the drive model presented in the paper.

## ACKNOWLEDGEMENT

The authors express their gratitude to Professors David Trumper and Steven Leeb at MIT for their valuable insights into the design of the drive and control systems.

## Appendix

The block diagram of the linearized model of the actuator and the nonlinear model of the mechanical dynamics of the actuator after elimination of the electrical dynamic using a high bandwidth are shown in Fig. 18.

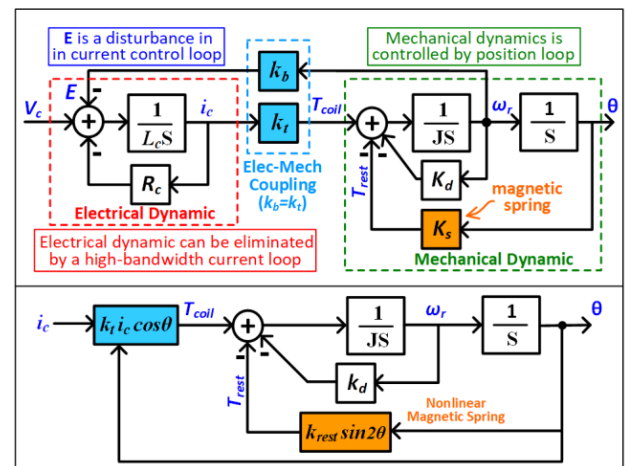

**Sajjad Mohammadi** (S'13) received the B.S. degree in electrical engineering from the Kermanshah University of Technology, Kermanshah, Iran, in 2011, the M.S. degree in electrical engineering from the Amirkabir University of Technology, Tehran, Iran, in 2014, and the M.S. and Ph.D. degrees in electrical engineering and computer science from the Massachusetts Institute of Technology (MIT), Cambridge, MA, USA, in 2019 and 2021, respectively. Now, he is a postdoctoral associate in precision motion control laboratory in the mechanical engineering department at MIT. His research interests include the design of electric machines, drives, power electronics, and power systems. He attained a number of awards as the 2022 George M. Sprowls Outstanding Ph.D. Thesis Award from MIT, 2014 Best MSc Thesis Award from IEEE Iran Section (nationwide), 2014 Best MSc Thesis Award from Amirkabir University , 2nd place in National Chem-E-Car Competitions 2009 held by Sharif University of Technology, Tehran, 5th place in International Chem-E-Car Competitions 2009 held by McGill University, Montreal, Canada, 2nd place in the Humanoid Motors together with 1st place in Technical Challenge in International Iran-Open 2010 Robotics Competitions, Tehran, 3rd place in National Chem-E-Car Competitions 2010 in Razi University, Kermanshah, and 1st place in the Humanoid Robot League together with the 1st place in Technical Challenge both in International AUTCUP 2010 Khwarizmi Robotics Competitions at Amirkabir University, Tehran.

**William R. Benner, Jr.** is president and CTO of Pangolin laser systems. As president, he sets the general strategic direction for the company and oversees all aspects of company operations. As CTO, he is in charge of all hardware and software development as well as research for new products and new directions for the laser display industry. In addition to having received more than 25 international awards for technical achievement, products invented by Benner and manufactured by Pangolin are currently used by some of the best-known entertainment and technology companies in the world, including Walt Disney World, Universal Studios, DreamWorks pictures, Boeing, Samsung, and Lawrence Livermore Labs. Benner holds more than 50 U.S. and International patents. He has articles published in the SMPTE Journal, The Laserist, EDN magazine, and Motorola's Embedded Connection magazine, and is co-author of the Amazon.com best-selling business books as well as a #1 best-selling book on Laser Scanners. Benner has also been featured on NBC, ABC, CBS, and FOX television affiliates. Beyond his work at Pangolin, Benner has also served on several boards as well as serving for seven years as ILDA's Technical Committee Chairman. He contributed expertise to outside companies, including Bliss Lights, Cambridge Technologies, RMB Miniature Bearings.

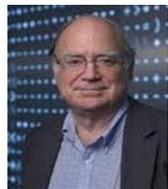

**James L. Kirtley, Jr.** (M'71–SM'80–F'91–LF'11) received the Ph.D. degree from the Massachusetts Institute of Technology (MIT), Cambridge, MA, USA, in 1971. Currently, he is a Professor of electrical engineering at MIT. He was with General Electric, Large Steam Turbine Generator Department, as an Electrical Engineer, for Satcon Technology Corporation as Vice President and General Manager of the Tech Center and as a Chief Scientist and as the Director. He was Gastdozent at the Swiss Federal Institute of Technology, Zurich (ETH), Switzerland. He is a specialist in electric machinery and electric power systems. Prof. Kirtley, Jr. served as Editor-in-Chief of the IEEE TRANSACTIONS ON ENERGY CONVERSION from 1998 to 2006 and continues to serve as an Editor for that journal and as a member of the Editorial Board of the journal Electric Power Components and Systems. He was awarded the IEEE Third Millennium medal in 2000 and the Nikola Tesla prize in 2002. He was elected to the United States National Academy of Engineering in 2007. He is a Registered Professional Engineer in Massachusetts, USA.

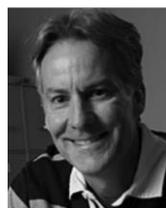

**Jeffrey H. Lang** (F'98) received his SB (1975), SM (1977) and PhD (1980) degrees from the Department of Electrical Engineering and Computer Science at the Massachusetts Institute of Technology. He joined the faculty of MIT in 1980 where he is now the Vitesse Professor of Electrical Engineering. He served as the Associate Director of the MIT Laboratory for Electromagnetic and Electronic Systems from 1991 to 2003, and as an Associate Director of the MIT Microsystems Technology Laboratories from 2012 to 2022. Professor Lang's research and teaching interests focus on the analysis, design and control of electromechanical systems with an emphasis on: rotating machinery; micro/nano-scale (MEMS/NEMS) sensors, actuators and energy converters; flexible structures; and the dual use of electromechanical actuators as motion and force sensors. He has written over 360 papers and holds 36 patents in the areas of electromechanics, MEMS, power electronics and applied control. He has been awarded 6 best-paper prizes from IEEE societies, has received two teaching awards from MIT, and was selected as an MIT MacVicar Fellow in 2022. He is a coauthor of Foundations of Analog and Digital Electronic Circuits published by Morgan Kaufman, and the editor of, and a contributor to, Multi-Wafer Rotating MEMS Machines: Turbines Generators and Engines published by Springer. Professor Lang is a Life Fellow of the IEEE, and a former Hertz Foundation Fellow. He served as an Associate Editor of Sensors and Actuators between 1991 and 1994. He has also served as the Technical Co-Chair and General Co-Chair of the 1992 and 1993 IEEE MEMS Workshops, respectively, and the General Co-Chair of the 2013 PowerMEMS Conference.